# Glassy Domain Wall Matter in KH$_2$PO$_4$ Crystal: Field-Induced Transition

Jitender Kumar and A.M. Awasthi[*]

Thermodynamics Laboratory, UGC-DAE Consortium for Scientific Research,

University Campus, Khandwa Road, Indore- 452 001, India

## ABSTRACT

We have investigated the domain wall (DW) dielectric response of potassium dihydrogen phosphate (KH$_2$PO$_4$) crystal under 0-500V dc-bias electric field. Activated DW-contribution onsets freezing at $T_f(\omega, V)$, some 27K below the ferroelectric $T_C$; timescale $\tau_f(T, V)$ exhibiting Vogel-Fulcher (VFT) divergence. Sharply distinct low- and high-field behaviors of $T_C(V)$, DW-$T_g(V)$, VFT-$T_0(V)$, barrier energy $U_a(V)$, and DW glass-fragility $m(V)$ signify a field-induced transition from randomly-pinned/vitreous to clustered/glass-ceramic phases of domain wall matter. Field-hysteresis ($\varepsilon'_{\text{poled}} > \varepsilon'_{\text{unpoled}}$) observed at high dc-bias indicates coexistent unclustered DW phase, quenched-in during the field-cooling. We construct a paradigm $T$-$E$ phase diagram depicting the complex glassy patterns of domain wall matter.

---

[*] Corresponding Author: amawasthi@csr.res.in

Potassium dihydrogen phosphate $KH_2PO_4$ (KDP) is an optical material widely used for frequency conversion and optoelectronic switching in laser systems.[1] KDP system has long been investigated for the structural phase transition responsible for its ferroelectricity,[2] and is a good example of a hydrogen-bonded ferroelectric. In the room temperature structure of KDP, hydrogen ions are statistically distributed over their two equilibrium positions. The two sites are about 0.4Å apart on the O-H-O bond. Below the Curie point, hydrogen in KDP is ordered, with two hydrogen atoms near every $PO_4$ group. KDP polarizes along the crystallographic *c*-axis and shows a ferroelectric transition at $T_C$ =123K. The crystal has anomalously high dielectric constant in the temperature range below $T_C$, compared to the predicted Landau theory roll-off. Below certain temperature $T_f$ ($< T_C$), permittivity falls back onto its expected phenomenological behavior. The excess $\varepsilon'$ over $T_f \leq T \leq T_C$ (the so-called plateau region) is believed to be due to the dynamics of the domain walls (DW's).[3-6] Abrupt decrease of dielectric permittivity below $T_f$ (DW freezing temperature) is thus attributed to the dynamical arrest of the domain walls.[6-10] Dielectric constant shows a kink here and imaginary permittivity peaks dispersively.[10-14] Here, we report the effects of DC-bias electric field on the dielectric response and the glassy DW kinetics. We characterize the glass-forming attributes of domain wall matter, as tuned by the applied bias-field. We also examine the effects on the dielectric response under poled (field-cooled, FC) and unpoled (zero field-cooled, ZFC) conditions. This work is useful as domain walls play an important role in the dielectric properties of ferroelectrics.

A suitable crystal was cut to the dimensions of 5.9x5.5mm & 2.2mm thick (along *c*-axis), and thin coatings of silver paint were deposited on both faces of the sample for proper electrode contact. We used Lakeshore 340 temperature controller, with the temperature stability better than ±0.02K, to scan the ferroelectric transition ($T_C$ = 123K) of the sample. For the dielectric measurements, we used the High Performance Frequency Analyzer (Alpha-A) and the High Voltage Booster (HVB) for dc-biasing (both from Novo Control). For the high-field dielectric work we designed/fabricated a dedicated probe for measurements over 350K to liquid-He range. Here we apply 1V ac signal to measure the dielectric response of the sample under various DC-bias fields. In our measurements on KDP, we have used the frequency range from 1Hz to 10kHz, and the applied dc-bias voltage is selected over 0 to 500V for all the temperatures. In the zero-field cooled (unpoled) and field-cooled (poled) measurements, we follow the standard protocol.

Figure 1a shows the real permittivity ($\varepsilon'$) vs. temperature at one of the probed frequencies for a number of applied dc-bias fields, with clear paraelectric (PE, tetragonal) to ferroelectric (FE, orthorhombic) transition at (zero-field) $T_C$ =123K. The dielectric constant is slightly lower for higher frequencies, and is well-matched with the earlier reports.[15-16] Rather high $\varepsilon'$ value (vis-à-vis its phenomenological expectation) below $T_C$ reflects the additional (domain-wall) contribution, that reduces with the applied bias field $E_{dc}$, mainly due to the enhanced DW-pinning.[16] Glassy $\alpha$-relaxation of DW's shows up in $\varepsilon''$ (fig.1b); their (dispersive) maxima at $T_f$ (< 100K, obtained as the peak-temperature) mark the freezing-onset of the activated DW dynamics.[7] Another sub-$T_C$ $\varepsilon''$-peak[16-17] is observed, whose peak-maximum frequency $\omega_m(V, 120K)$ we identify with a metric of the DW pinning-frequency $\omega_{DW}^{pin}$.

At a selected temperature 120K [< $T_C$(500V)], to cover the DW-character at all the bias-fields), fig.1b top-inset displays mild-to-steep changeover in $E_{dc}$-dependence of the electro-capacitance $\{\varepsilon'(V)/\varepsilon'(0)-1\}$ at 10kHz frequency (found similar at all frequencies). Moreover, in fig.1b bottom-inset, the 120K loss-spectra (peak-fitted) at various $E_{dc}$'s show a turnaround in the "DW pinning-frequency" $\omega_m(V, 120\mathrm{K}) \propto \omega_{\mathrm{DW}}^{pin}$, with $\omega_m$ obtained from the 120K peak-maxima.[17] These unmodelled-results signify a critical $V_{cr}$ =150V ($E_{cr} \approx 70\mathrm{kV/m}$), separating different DW-configurations. While up to $V_{cr}$, increased local-pinning of the DWs is witnessed as expected, over the ($\leq T_C(E)$, $\geq E_{cr}$) regime DW's organize into finite correlation-length clusters (DWC).[16] Besides heralding a demise of the individual DW-contribution to the complex permittivity, the DWC manifest low-energy collective excitations. Decreasing "pinning-frequency" and the stronger $E_{dc}$-dependent electro-capacitance amply evidence this qualitative change in the DW degrees of freedom. Now, the translational motion typical of individual DW gives way to the overdamped dynamics of bulkier DWC; one expects their dissipative rocking/vibratory/breathing-modes response to contribute to the ac-permittivity. Immediately below $T_C$, highly-suppressed contribution to $\varepsilon'$ and spectrally-broader loss-peak at reduced 'pinning' frequency now characterize the dominantly relaxational-attribute of the DWC, determined by their size-scale.

We now investigate the effects of the dc-bias field on the DW-relaxation freezing process. This is clearly shown in $\varepsilon''$ to be of kinetic nature; on increasing the applied $E_{dc}$, the loss-peak temperature ($T_f$) shifts downwards (fig.2). Insets show the "peak-contributions" $\varepsilon'_{\mathrm{DW}}$ of domain-walls to the real permittivity $\varepsilon'$, obtained by subtracting a supposedly 'domains-only' background (polynomial-interpolated by joining the measured $\varepsilon'$-data, at $< T_C$ and that $<< T_f$). Notice the continued dispersion in $\varepsilon'_{\mathrm{DW}}$ right up to $V_{dc}$ = 500V.

DW-freezing kinetics is analyzed in terms of $\tau_f(T, V)$, read off the $\varepsilon''_v(\omega, T)$ loss-peak maxima. The Arrot-plot iso-potentials of $\ln(\tau)$ vs. $1/T$ are shown in fig.3. The DW relaxation comes across as the stretched-exponential type;[7] the effective times associated with the process fit the Vogel-Fulcher behavior.[18-19]

$$\tau = \tau_0 \exp\left(\frac{U_a}{T-T_0}\right) \qquad (1)$$

Here, $T_0$ is the Vogel-Fulcher temperature ($\approx T_K$, the Kauzmann temperature[20]) and $U_a = DT_0$ is the barrier-energy for thermally activated process ($\equiv k_B DT_0$ in Joules). $D$ is identified as the glass strength[21-23] (a strong glass retains its glassy character over time and/or against external influence). While $T_0(V)$ ($\sim$ 89K) mildly humps (reported $T_0(0)$ $\sim$ 70K),[16] unusually low $U_a$ ($\sim$10K $<<$ $T_g$) here is much susceptible to the applied dc-bias field (main inset). Here, we could fit the low- and high-field activation energies to different logarithmic dependences [$U_a = A - B\ln(E_{dc} + C)$]; the sharp break in behavior being explicit in the lin-log plot (sub-inset). The two fits sharply "switchover" at $E_{cr} =$ 150kV/2.2m (meant as the critical voltage of $V_{cr} =$ 150V across 2.2mm-thick specimen), precisely identified earlier as marking the steeper drop-down in $\varepsilon'(V,$ 120K) and the turnaround of $\varepsilon''(V, 120\text{K})$-peak-frequency vs. $E_{dc}$. We assert these dramatic manifestations as but consequent to the *transition-point* status of $E_{cr}$. Furthermore, the timescale-divergence of 180°-DWs is *unlike* the polar nanoregions (PNR) in relaxors, as both the $\varepsilon'_{\text{DW}}$ and $\varepsilon''$ peak-contributions (fig.2) are dispersive up to the highest bias-fields used. Contrarily, the relaxor PNR's merge under high dc-field to give rise to robust (non-dispersive/lossless) long-range ferroelectricity.[24-25]

For a comparative/unified study of domain-wall-matter in ferroelectrics, a paradigm template emerges from the rich features manifested here. To this end, we show in fig.4 a DW phase diagram construction, as observed in the KDP crystal. With a discontinuity at $E_{cr}$, the FE-$T_C(V)$ marks the emergence of domain-wall-matter (DWM). Compared to $T_g$ ($\approx T_f$ at low $\omega$'s) with relatively 'large' drop (mainly above $E_{cr}$), the close-by $T_0$ with a hump-structure serves as the non-trivial demise-phase-boundary for the DWM. Bias-field squeezing the narrow sluggish-DW window (inset, left-scale) explores this very rare critical glass-regime; inflexion point at $E_{cr}$ delineating the two DW-phases. The locally-pinned (clustered) DW-regimes are evident in the rise (fall) of $\omega_m(V, 120\mathrm{K}) \propto \omega_{\mathrm{DW}}^{pin}$ below (above) $E_{cr}$ (inset, right-scale).

The kinetic glass-fragility[23, 26-27] $m = \dfrac{d \log \tau}{d\left(T_g / T\right)}\bigg|_{T=T_g} = \left(\dfrac{D}{\ln 10}\right) \times \left(\dfrac{T_0}{T_g}\right) \times \left(1 - \dfrac{T_0}{T_g}\right)^{-2}$ (main

panel, right-axis) combines a thinning sluggish-zone-width [$(T_g\text{-}T_0) \to 0$ favoring the fragility-increase[23]] with rather low & steeper-falling energy-barrier ($U_a = DT_0 << T_g$) in slightly *raising* the DW-vitreousity up to $E_{cr}$. Further up however, DW-clustering phenomenon (*c.f.*, micro-crystallization) reverts this trend; the DW-matter acquires a rather fragile ($m \to 600$) glass-former (GF) attribute under high bias-fields. By all accounts, the critical $E_{cr} \approx 70\mathrm{kV/m}$ registers a transition from randomly-pinned (rather strong GF) to clustered (super-fragile GF) phase of DWM.

To further explore the nature of field-induced phase transition (FIPT) of DWM, we compare (fig.5) our in-field (500V) warm-up permittivity data obtained in unpoled (zero-field-cooled, $\varepsilon'_{\mathrm{ZFCW}}$) and poled (field-cooled, $\varepsilon'_{\mathrm{FCW}}$) runs, against the virgin (zero-field-cooled and zero-field-warmed $\varepsilon'_{\mathrm{virgin}}$) data background. In close view, below $T_C(V)$ down to a temperature $T_{cl}(V, \omega)$, the two *in-field* permittivity data are found the same to within the noise level, suggesting $T_{cl}$ as the "clustering" temperature. Further, we confirm the splitting ($\varepsilon'_{\mathrm{virgin}} > \varepsilon'_{\mathrm{FCW}} > \varepsilon'_{\mathrm{ZFCW}}$) over $T_f(V, \omega) \le T \le T_{cl}(V, \omega)$ at all frequencies (fig.5b inset). Higher permittivity in the poled case here is compelled by a kinetic-coexistence (*frequency*-dependent split [$\varepsilon'_{\mathrm{FCW}}(\omega) - \varepsilon'_{\mathrm{ZFCW}}(\omega)$], fig.5b inset, with $T_{cl}^{lo-\omega} < T_{cl}^{hi-\omega}$) of clustered (stable/transformed) *and* pinned (metastable/field-quenched) DW phases. We invoke the resemblance of this "field-hysteresis" in permittivity with with that of the order-parameter (e.g., magnetization) in the disorder-broadened first-order phase transition (FOPT,[28] e.g., in magnetism). Therefore, in the present case the split may be a signature of the first-order character of the field-induced phase transition (FIPT).

In conclusion, our dielectric study of KDP crystal below FE-$T_C$ reveals that glassy domain-wall-matter (DWM) is sizescale-organized under high dc-bias-fields. Vogel-Fulcher kinetics of domain-wall freezing ($\alpha$-relaxation loss-peak at $T_f$) provides mildly $E_{dc}$-variant (humped) VFT-temperature ($T_0 \sim 90$K) and unusually smaller (vis-à-vis both $T_0$ and $T_f$) activation energy $U_a$ (~10K), that starkly delineates the low- & high-field regimes as well. At high bias-fields, anomalous downshift in a metric of DW-pinning-frequency, non-trivial changeover of sub-$T_C$ electro-capacitance, and contraction of sluggish-DW $T$-regime are all traceable to the clustering of the domain walls. Higher poled-permittivity (vs. unpoled one) obtained is compatible with the coexistence of pinned (metastable/field-quenched) and clustered DWM phases. Our $T$-$E$ phase diagram correlates abrupt changes across an $E_{cr} \sim 70$kV/m in $T_C(V)$, $T_g(V)$, $T_0(V)$, $U_a(V)$, and glass-fragility $m(V)$ parameters, characterizing the mobile/sluggish (above/below $T_g$) states of DWM. Our maiden findings mandate a field-induced transition between locally-pinned (rather strong glass-former) and clustered (super-fragile glass-former) phases of domain-wall-matter.

**Acknowledgements**

We thankfully acknowledge receiving the KDP crystals from Raja Ramanna Center for Advanced Technology, Indore. J.K. appreciates help provided by S.

Bhardwaj to modify the dielectric probe for high-voltage applications. Dr. P. Chaddah is thanked for discussions on FOPT and for his scientific support and encouragement.

**Figure Captions**

**Figure 1.** Real (a) and imaginary (b) parts of permittivity vs. temperature for the KDP single crystal investigated at various dc-bias electric fields ($V_{dc}$ = 0-500V). The field-variations in $\varepsilon'$ and in $T_C$ are most clearly visible at 1Hz. On the other hand, the two sets of relaxation peaks (around $T_f$ and just below $T_C$) are most distinct/separate in the $\varepsilon''$ at 1kHz. The slowly decreasing plateau in $\varepsilon'$ below $T_C$ reflects the excess contribution due to the domain walls (DW's), whose freezing is marked by the field-dependent relaxation peaks in $\varepsilon''$ about $T_f \sim$ 96K. Inset in (a) shows the systematic decrease in $T_C(V)$ and $\varepsilon'(T_C^V)$. Insets in (b) show the electro-capacitance (10kHz) and the relaxation-spectra [both at 120K < $T_C^{500V}$]; respectively depicting steeper drop-down (upper inset) and turnback of the peak-frequency (lower inset) beyond a $V_{cr}$ = 150V ($\equiv E_{cr} \sim$ 70kV/m). Bias-field is in units of Volts per 2.2mm-thick specimen.

**Figure 2.** Dispersion of the glass-relaxation peak ($\varepsilon''$) at (a) zero dc-bias field and at (b) $V_{dc}$ = 500V ($\equiv E_{dc} \sim$ 225kV/m). Estimates of the corresponding $\varepsilon'_{DW}$-contributions are shown in the insets, obtained by subtracting a (polynomial-fitted) supposed Landau phenomenological background from the measured $\varepsilon'$.

**Figure 3.** Arrot-plot iso-potentials ($\ln \tau$ vs. $T^{-1}$) of the glass-relaxation time in the DW freezing regime. Inset shows the barrier activation energy vs. the applied field, obtained from the Vogel-Fulcher fits to the main panel curves. The lin-log plot (sub-inset) sharply delineates at $V_{cr}$ = 150V ($\equiv E_{cr} \sim$ 70kV/m), the two different logarithmic dependences found for $U_a(E_{dc})$ in the low- and high-field regimes. The Vogel-Fulcher temperature $T_0 \sim$ 89K is found as nominally $E_{dc}$-dependent (see fig.4). Bias-field is in units of Volts per 2.2mm-thick specimen.

**Figure 4.** Paradigmatic phase diagram construction collates the various kinetically-active and dormant regimes of glassy domain wall matter (DWM) and their phase-boundaries. Inset emphasizes the special bias-field ($E_{cr} \sim 70$kV/m) as separating the locally-pinned and micro-clustered phases of DWM. Overwhelming evidence here from both measured and derived results clearly mandates a field-induced order-disorder transition at $E_{cr}$ between phases of DWM; having distinct glass-fragility characters referred to their locally-pinned-disorder vs. clustered-organization. Bias-field is in units of Volts per 2.2mm-thick specimen.

**Figure 5.** Comparison of the (500V) unpoled (zero-field-cooled, ZFC), poled (field-cooled, FC), and the virgin (0V) permittivity, all taken during the warm-up. Noticeably, poled and unpoled data are split over $T_f(V, \omega) \leq T \leq T_{cl}(V, \omega)$; the clustered-phase emerges only at a temperature $T_{cl}(V) < T_C(V)$ in the poled case, and also seems to disappear above $T_{cl}(V)$ in the unpoled case. Split of the two in-field permittivity-data is due to the larger differential contribution of the quenched-in metastable pinned-phase (data below $E_{cr}$ in fig.1b upper inset), coexistent with the (balance/unquenched and transformed) stable clustered phase of domain wall matter.

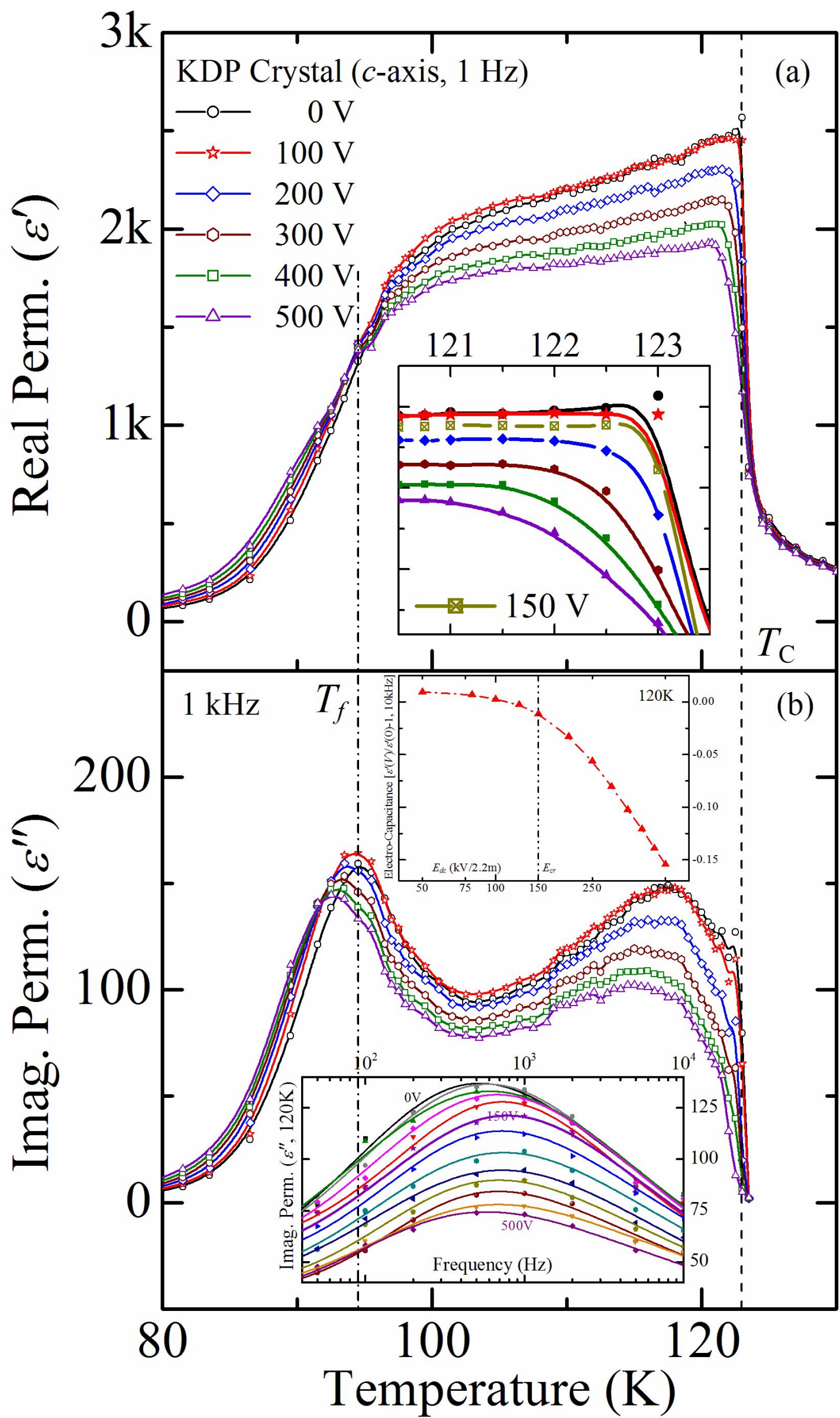

**(a)**

KDP Crystal (*c*-axis, 1 Hz)

- 0 V
- 100 V
- 200 V
- 300 V
- 400 V
- 500 V

Real Perm. ($\varepsilon'$)

3k
2k
1k
0

121   122   123

150 V

$T_C$

**(b)**

1 kHz

$T_f$

Imag. Perm. ($\varepsilon''$)

200

100

0

Electro-Capacitance [ε'(V)-ε'(0)×1.10kHz]   120K

$E_{dc}$ (kV/2.2m)   $E_p$

0.00
−0.05
−0.10
−0.15

50   75   100   125   150

Imag. Perm. ($\varepsilon''$, 120K)

0V
150V
500V

Frequency (Hz)

$10^2$   $10^3$   $10^4$

125
100
75
50

80   100   120

Temperature (K)

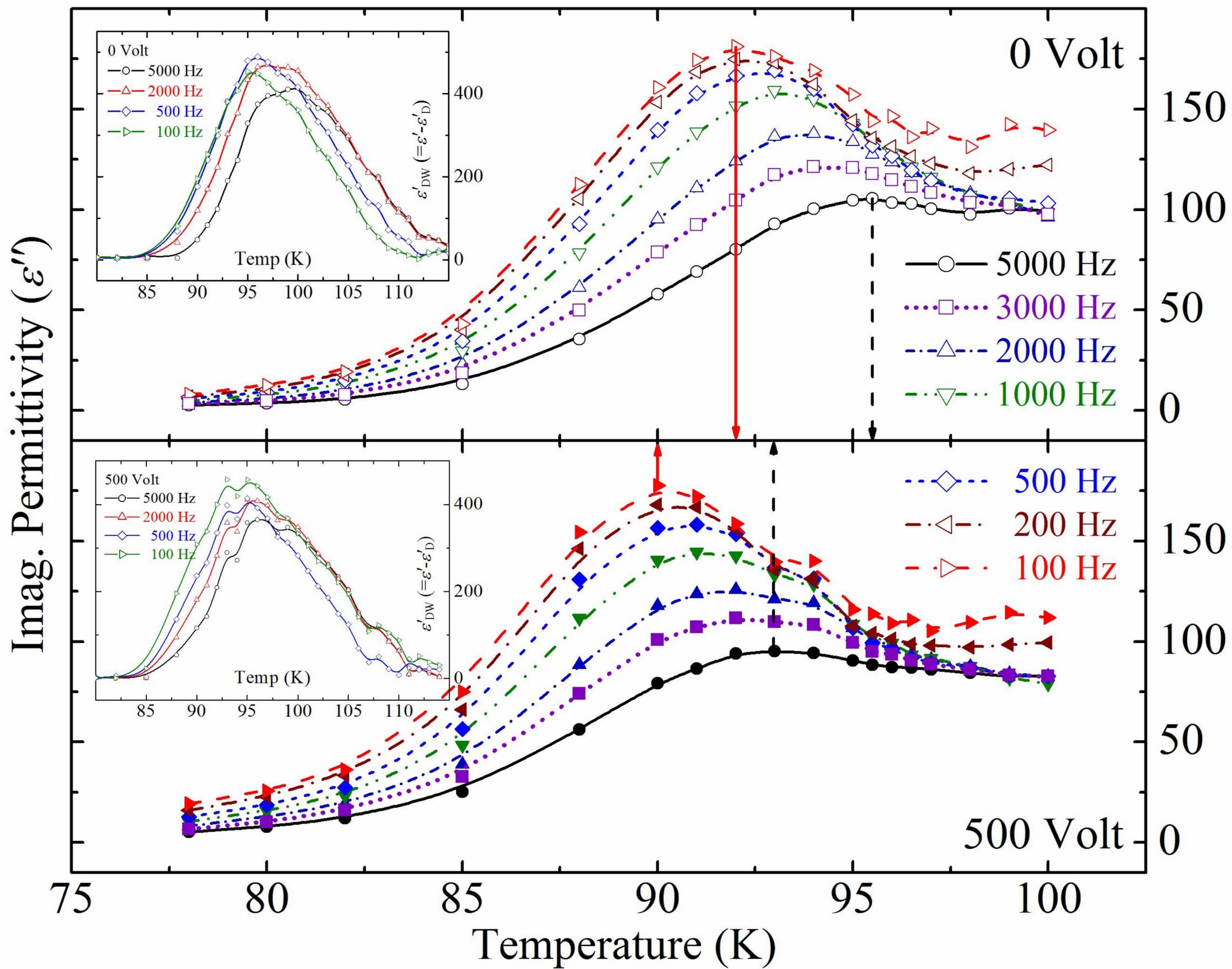

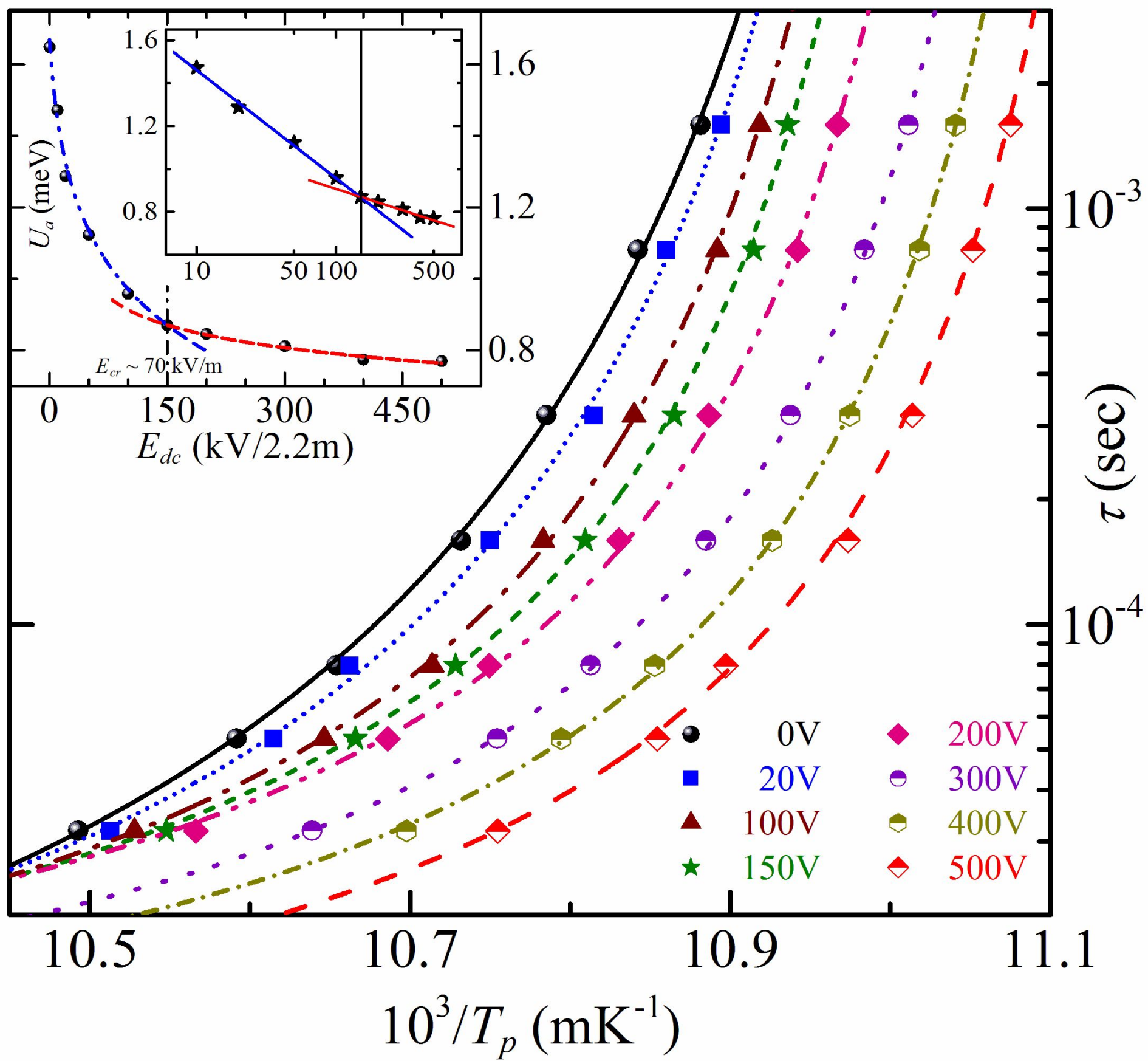

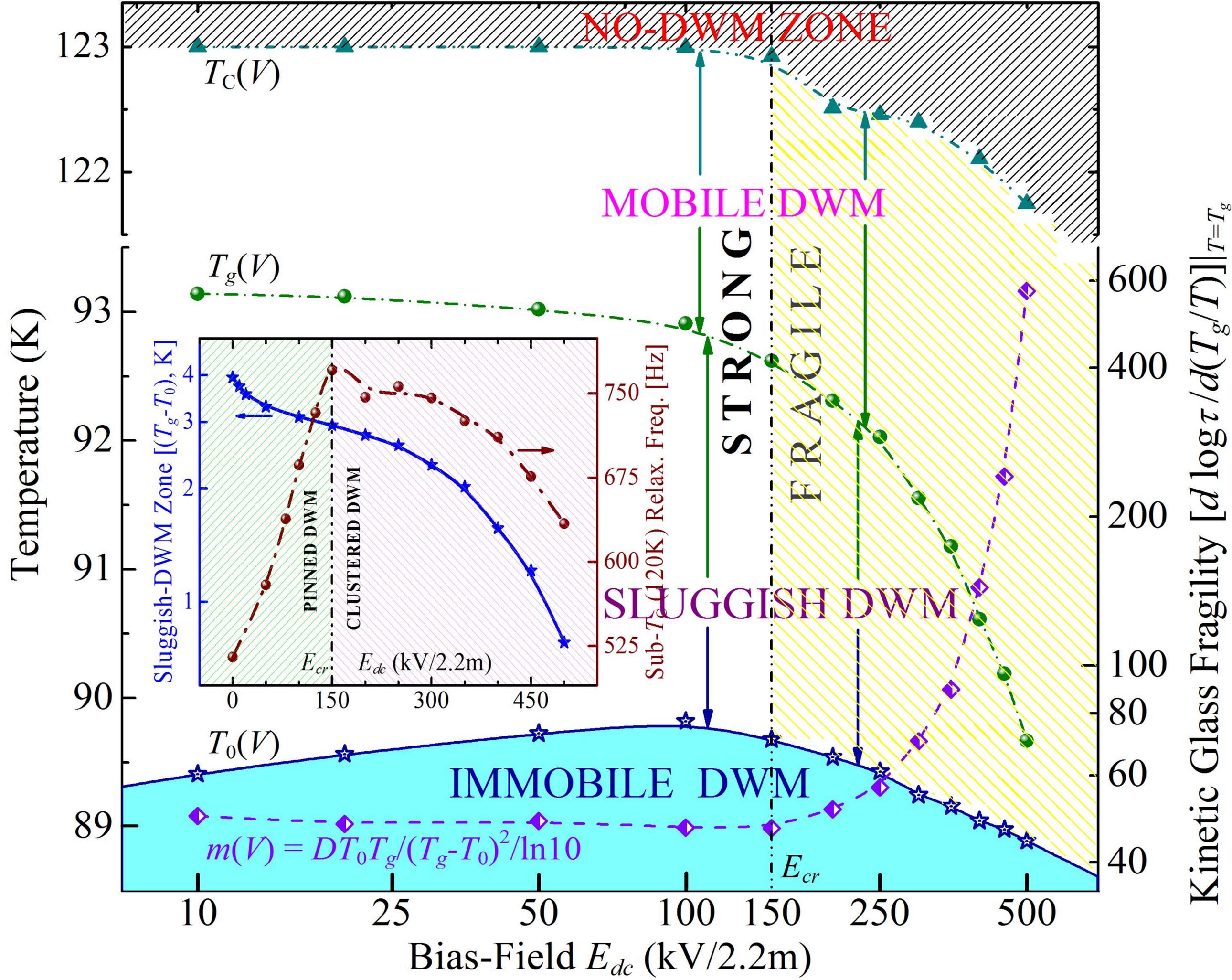

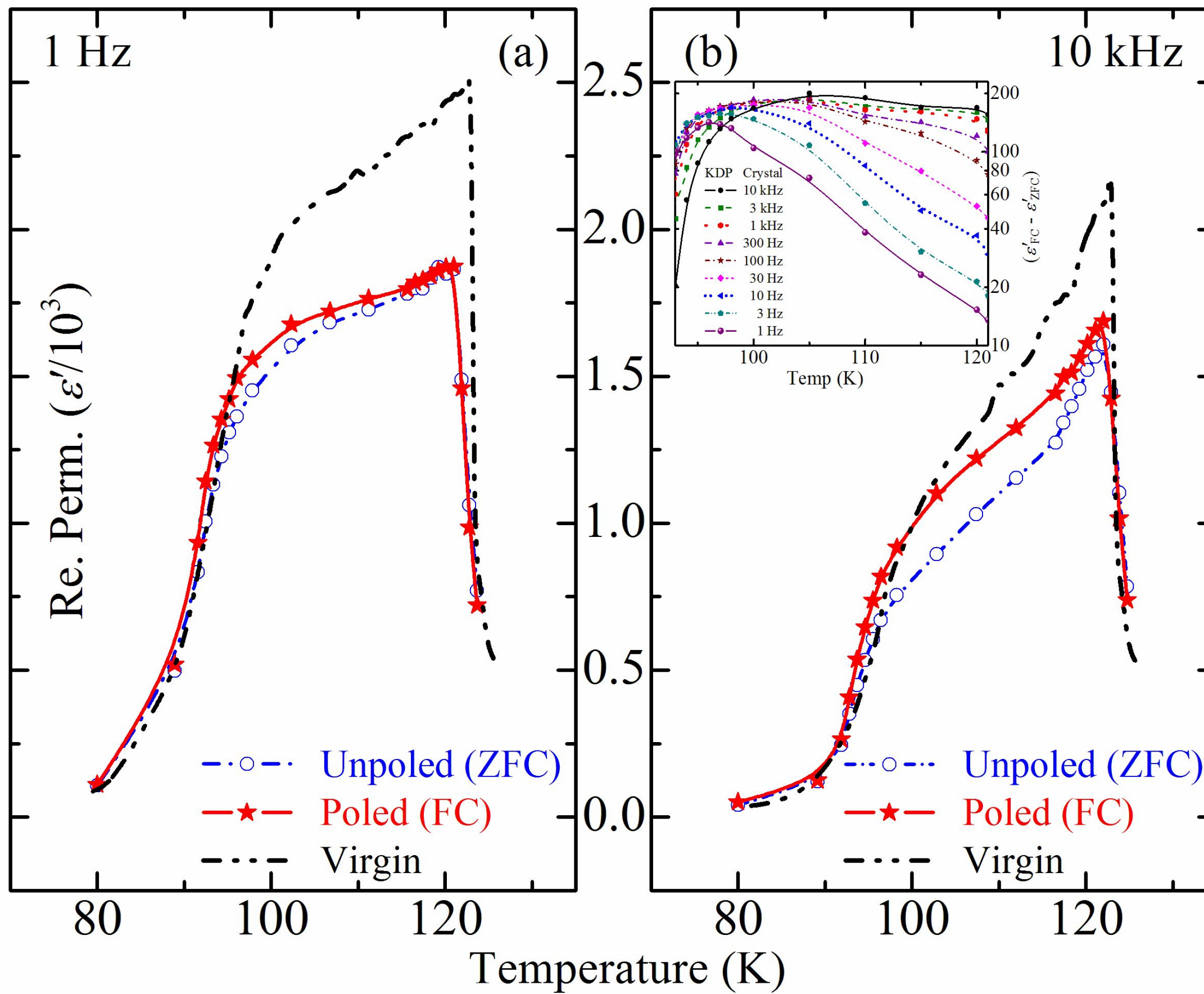